\documentclass{appolb}

\usepackage{graphicx}
\usepackage{subfigure}
\usepackage{cite}

\begin{document}

\title{Hadron resonances, large $N_c$, and the half-width rule\thanks{Presented by ERA at {\em Light
      Cone 2012}, Cracow, 8-13 July 2012}
\thanks{
    Supported in part by the Polish Ministry of Science and Higher
    Education grant N~N202~263438, Polish National Science Center
    grant DEC-2011/01/B/ST2/03915, Spanish DGI and FEDER funds with
    grant FIS2011-24149, Junta de Andaluc{\'\i}a grant FQM225 and the
    german DFG through the Collaborative Research Center “The
    Low-Energy Frontier of the Standard Model” (SFB 1044).}
}
\author{Enrique Ruiz Arriola \address{Departamento de
    F\'{\i}sica At\'omica, Molecular y Nuclear \\ and Instituto Carlos
    I de F{\'\i}sica Te\'orica y Computacional, Universidad de
    Granada, E-18071 Granada, Spain} \and Wojciech Broniowski
  \address{The H. Niewodnicza\'nski Institute of Nuclear Physics PAN,
    PL-31342 Krak\'ow \\ and Institute of Physics, Jan Kochanowski
    University, PL-25406 Kielce, Poland } \and Pere Masjuan
  \address{Institut f\"ur Kernphysik, Johannes Gutenberg
    Universit\"at, Mainz D-55099, Germany} } \date{25 October 2012}

\maketitle
\begin{abstract}
We suggest using the half-width rule to make an estimate of the
$1/N_c$ errors in hadronic models containing resonances. We show
simple consequences ranging from the analysis of meson Regge
trajectories, the hadron resonance gas at finite temperature and
generalized hadronic form factors.
\end{abstract}

 
\bigskip

\section{Introduction}

The Particle Data Group (PDG) tables~\cite{Nakamura:2010zzi}
represent the nowadays consensus of the particle spectrum, and it is
quite legitimate to ask the question on the completeness or redundancy
of the states listed there. The quantum numbers accommodated by the
quark model for the u,d,s flavored mesons, $n ^{2S+1}L_J$ furnishes, till
now, a complete commuting set of observables. On the other hand, since
most of these states are unstable, a thorough understanding of the
physics summarized by the PDG is related to the concept of a
resonance.

However, the hadronic resonances are never observed directly but through
their decay channels, and the corresponding cross section also depends
on the particular production process. The standard and unique quantum
mechanical definition of a resonance is via a pole in the second Riemann
sheet of the complex-$s$ plane in a scattering amplitude containing
such a resonance, although the (complex) residue depends on the
process.  This definition has the advantage of being quite universal
regarding the pole position, but can only be applied if the amplitude
can be analytically continued in a reliable way.  Indeed, complex
energies cannot be measured experimentally nor simulated by lattice QCD
calculations, and basically an extrapolation is needed; a potentially 
uncontrolled arbitrary procedure~\cite{Ciulli:1975sm}. Thus, in
order to deduce these poles reliably, one must either have narrow
resonances, small backgrounds, or accurate amplitudes, requirements
which are rarely met in the PDG compilation~\cite{Nakamura:2010zzi}.

There are other and more handy definitions which apply to the physical
and real energy, such as the maximum in the speed plot, the time delay, or
the popular Breit-Wigner definition. While all these definitions should naturally
merge in the limit of narrow resonances, the finite widths build
systematic differences which introduce some inherent dependence on the
background. The upshot of the present discussion is that one should be
concerned with i)~what is the right value to quote and ii)~to what
confidence level can different values be considered as compatible. Again,
the PDG compilation incorporates different processes which quite often rely
on models or parameterizations.

\section{Large $N_c$ and the half-width rule}

A useful observation is that in the large $N_c$
limit~\cite{'tHooft:1973jz,Witten:1979kh} one has $\Gamma/M = {\cal O}
(N_c^{-1})$ and one finds~\cite{Arriola:2011en}
$\Gamma/M=0.12(8)$ both for mesons and baryons composed of the light u,d,s
quarks and listed by the PDG~\cite{Nakamura:2010zzi}. Most
mesonic and baryonic resonances stem from the $\bar q q
$ and $qqq$ bound states which become unstable once they are allowed
to decay in the continuum. We suggest that the maximum level of
discrepancy in quoting resonance mass parameters should just be compatible
with its own width, namely the interval $M_R \pm \Gamma_R/2$.

A model-independent way of looking at resonances in QCD is by
considering the two-point correlation functions.  Actually, in the
quenched approximation one may treat them as standard bound states.
Consider for instance the case of the $\rho$-meson, which is obtained
as a $\bar q q$ state from the vector-vector correlation function.
The Lehman representation of the resonance two-point function is
\begin{eqnarray}
D(s)= \int_0^\infty d \mu^2 \frac{\rho(\mu^2)}{\mu^2-s-i {0^+}} ,
\end{eqnarray}
suggesting a probabilistic interpretation of the line shape
\begin{eqnarray}
P(\mu)= Z \rho(\mu) 
\end{eqnarray}
as a function of the mass $\mu$. For a Breit-Wigner shape we
have
\begin{eqnarray}
  D_{\rm BW }(\mu)= \frac{1}{\mu^2-M^2-i \Gamma \mu} \to P_{\rm BW}(\mu)= \frac1{\pi}\frac{2 \Gamma \mu^2}{(\mu^2-M^2)^2 +\Gamma^2 \mu^2)} .
\end{eqnarray}
The random implementation for a given distribution is obtained by inverting 
the relation 
\begin{eqnarray}
P(\mu) d\mu  = dz 
\end{eqnarray}
with $z \in U[0,1]$ denoting a uniformly distributed
variable~\footnote{One can also use Gaussian variables for $\mu$ which
  have shorter tails.}. The half-width rule (HWR) consists of treating
the resonance mass as a random variable and propagating its effect to
all observables

\section{Mesonic Regge Trajectories}

Long ago it was suggested~\cite{Kang:1975cq} that linear confinement
in quark models implies the mesons {\it radial} Regge trajectories of
the form generalizing the Chew-Frautchi plots for the
angular momentum~\cite{RevModPhys.34.394}. The analysis of
Ref.~\cite{Anisovich:2000kx} of $M_{n}^2\,= M_0^2 + \mu^2 n$ gave
$\mu^2=1.25(15)$~GeV$^2$. The HWR amounts to
minimize~\cite{RuizArriola:2010fj,Masjuan:2012gc}
\begin{eqnarray}
\chi^2 = \sum_n \left(\frac{M_{n}^2-M_{n, {\rm exp}}^2}{\Gamma_{n} M_n} \right)^2 \, ,
\label{eq:chi2}
\end{eqnarray}
The construction of these trajectories requires a
choice on the possible quantum-number assignments. Following 
~\cite{Masjuan:2012gc}, the global result for the non-strange mesons 
was found 
\begin{equation} 
M^2= 1.38(4)n + 1.12(4) J -1.25(4)\, . \nonumber 
\end{equation}
Several models~\cite{Afonin:2006wt,Glozman:2007ek}, including
holographic approaches~\cite{deTeramond:2008ht}, assume a universal
radial and angular momentum slope, i.e., an exact $ (n+J) $-dependence.
As we can see, there seems to be a significant deviation from this
universality, similarly as in the relativistic quark
model~\cite{Ebert:2009ub}. 

A global fit which does not require a selection of states, but assumes
$\bar q q$ completeness, considers the staircase function for the
mesons, defined as
\begin{eqnarray}
N_{\rm mesons} (M)= \sum_{nLSJ f,f'} (2S+1) \Theta (M^2 - a n - b J - c_{f,\bar f'})
\end{eqnarray} 
where the spin-orbit and tensor force effects are neglected. Within
the mass range $0.5 {\rm GeV} \le M \le 1.85 {\rm GeV}$ and using
$\Delta M=10{\rm MeV}$ bins, it yields a flavor-dependent log-fit with
$a= 1.40 {\rm GeV}^2 $, $b=1.10 {\rm GeV}^2 $ $c_{n \bar n}=-1.23 {\rm
  GeV}^2 $, $c_{n \bar s}= c_{s \bar n}=-0.40 {\rm GeV}^2$ and $,c_{s
  \bar s}=-0.78 {\rm GeV}^2 $, in good agreement with the non-strange
single state determination and despite the fact that the degeneracy
plays a role in the fit.

\section{Hadronic density of states}

\begin{figure}[tb]
\begin{center}
\subfigure{\includegraphics[width=0.48\textwidth]{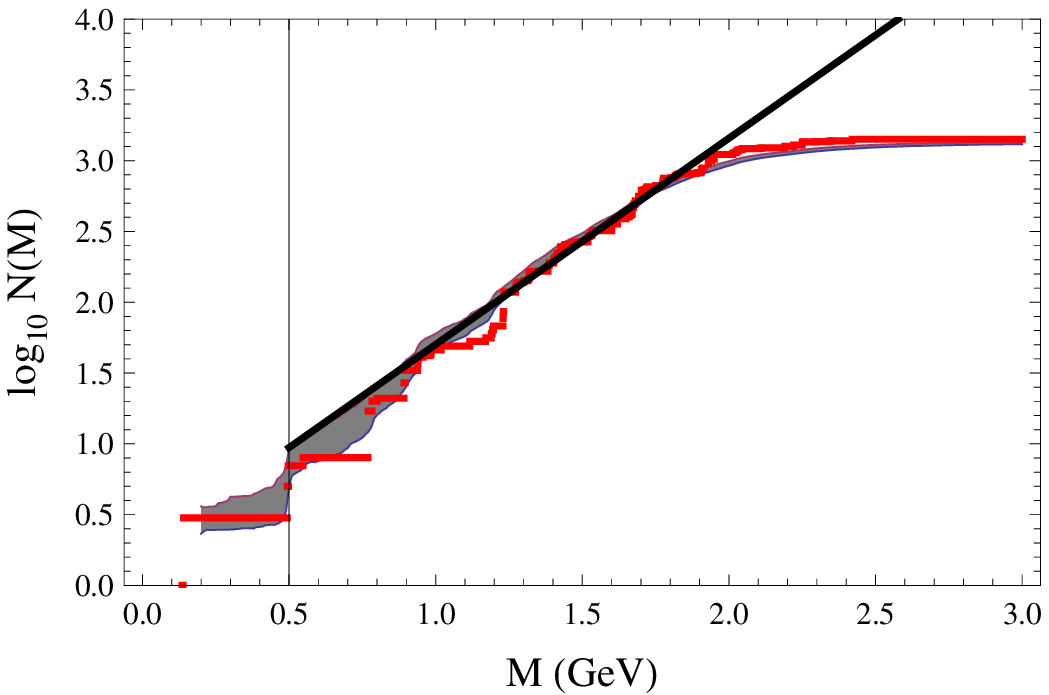}} \hfill
\subfigure{\includegraphics[width=0.48\textwidth]{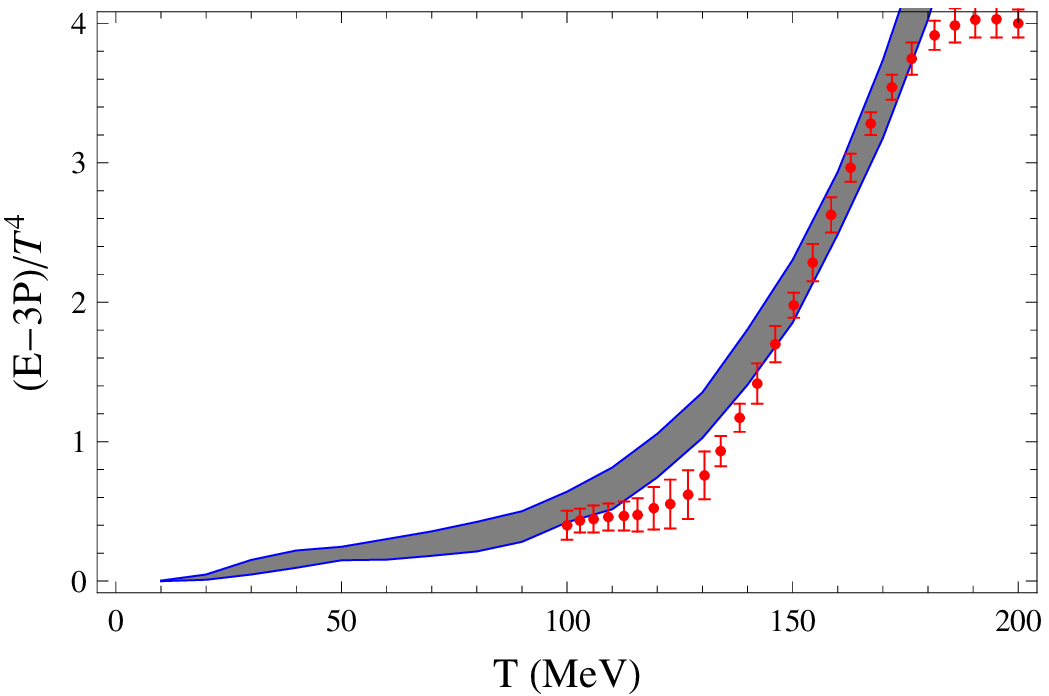}}
\end{center}
\vspace{-3mm}
\caption{Left: Cumulative number of hadrons from the
  PDG~\cite{Nakamura:2010zzi} when resonances are represented as
  random BW variables $M_R \pm \Gamma_R/2$ as a function of the
  energy. We also plot the exponential spectrum $N(M)=A e^{M/T_H}$
  with the Hagedorn temperature given by $T_H= 300(75) {\rm MeV}$.  
  Right: Trace
  anomaly of the Hadron Resonance Gas implementing the half-width rule
  compared to the lattice data of Ref.~\cite{Borsanyi:2010bp}
\label{fig:hw-hag}}
\end{figure}

A further direct application of the HWR concerns the
analysis of the cumulative hadron number
\begin{eqnarray}
N(M) = \sum_i g_i \Theta(M-M_i) 
\end{eqnarray}
where $M_i$ are the individual hadronic masses and $g_i$ are the
corresponding spin degeneracies (particles and antiparticles are
counted separately).  This yields a staircase function which is
presented in Fig.~\ref{fig:hw-hag} when all the PDG
hadrons~\cite{Nakamura:2010zzi} with the light $u,d,s$ quarks are
included.  The exponential growth of $N(M) \sim A e^{M/T_H}$ can be
distinctly seen, although, as pointed out in
Ref.~(\cite{Broniowski:2000bj,Broniowski:2004yh,Cohen:2011cr}), a
pre-exponent power cannot be extracted from the data. As seen in
Fig.~\ref{fig:hw-hag} the effect of taking into account the resonance
uncertainty naturally smooths the data and provides an estimate of the
finite width corrections, hence allowing an error analysis. Using the
BW distribution for the widths we get $T_H= 300(75) {\rm MeV}$ and
$A=1.758(2)$ with a $\chi^2/{\rm d.o.f.}  = 0.92$ in the range $0.5
{\rm GeV} \le M \le 1.85 {\rm GeV}$.

Likewise, the trace anomaly in QCD has been calculated on the
lattice~\cite{Borsanyi:2010bp} and below the cross-over transition to
the quark-gluon plasma it can be represented in the Hadron Resonance
Gas as follows
\begin{eqnarray}
\Delta=\frac{\epsilon-3p}{T^4} = \frac1{T^4} \int_0^\infty dM \frac{dN(M)}{dM} 
\int \frac{d^3
  k}{(2\pi)^3} \frac{(E_k-\vec k \cdot \nabla_k E_k )}{e^{E_k/T}\pm 1}
\end{eqnarray}
where $E_k=\sqrt{k^2+M^2}$ and $\pm$ corresponds to
Fermions/Bosons. As seen in Fig.~\ref{fig:hw-hag}, the half-width rule
provides an error estimate for $\Delta$ in the HRG which compares
favorably with the lattice data of the Wuppertal
group~\cite{Borsanyi:2010bp}.

\begin{figure}[tb] 
\begin{center}
\subfigure{\includegraphics[width=0.48\textwidth]{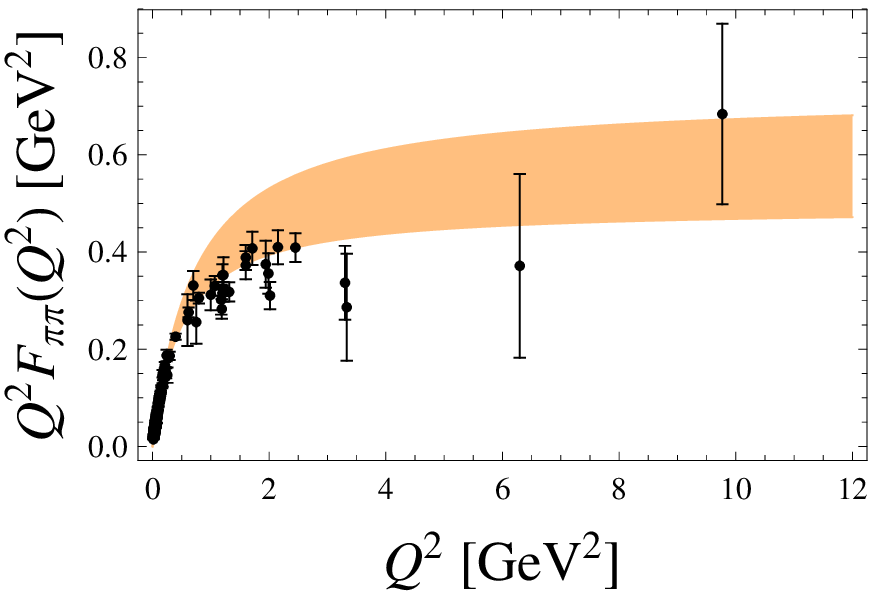}} \hfill
\subfigure{\includegraphics[width=0.48\textwidth]{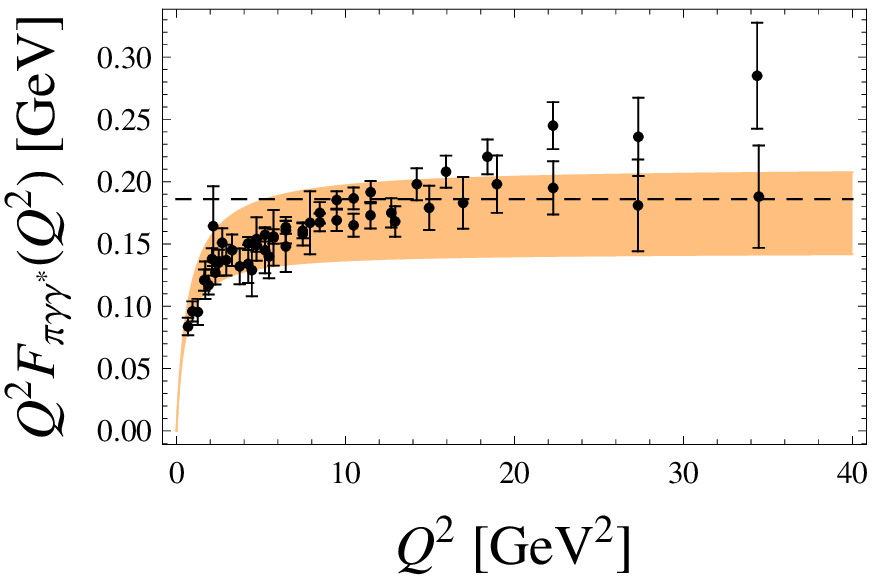}}
\end{center}
\vspace{-3mm}
\caption{The monopole form factors $F(Q^2)=m_\rho^2/(m_\rho^2+Q^2)$
  gaussian sampled with the HWR, $\pm \Gamma_\rho/2=\pm 0.75 {\rm
    MeV}$. Left: Charge pion form factor, $F_{\gamma^*
    \pi\pi}(Q^2)=F(Q^2)$. Right: Transition form factor, $F_{\pi
    \gamma \gamma^*}(Q^2)= F(Q^2) /(4\pi f_\pi)$
\label{fig:formfactors}}
\end{figure}

\section{Hadronic form factors}

As a final example of the HWR let us consider hadronic generalized
form factors as analyzed recently~\cite{Masjuan:2012sk}. 
The well-known fact is that in the large-$N_c$ limit of QCD
the generalized hadronic form factors, probing bilinear $\bar q q $
operators with given $J^{PC}$ quantum numbers, feature generalized
meson dominance of $\bar q q $ states with the same quantum
numbers,
\begin{eqnarray}
\langle A(p') | J (0) | \langle B(p) \rangle \sim 
\sum_{n} \frac{ c_n^{AB} m_n^2}{m_n^2-t},
\end{eqnarray}
where $m_n$ are the meson masses and $c_n^{AB}$ the suitable
couplings.  Thus generalized form factors at some finite momentum
transfer essentially measure the masses of the lowest lying mesons in
the given channel.  Saturating with the minimum number of mesons to
yield the known high $Q^2$ and applying the HWR produces an error band
estimate which we show in Fig.~(\ref{fig:formfactors}) for the pion
electromagnetic and transition form factors. Refinements and further
pion and nucleon form factors are presented in~\cite{Masjuan:2012sk}.
While large-$N_c$ behavior of hadronic quantities provides a unique
fingerprint of QCD. Quark-Hadron Duality allows to sidestep the
difficult problems by imposing the short-distance constraints.  Using
a minimum number of resonances leads to a sensible parameter
reduction.  Errors based on the half-width rule provide a reasonable
and large $N_c$ motivated estimate on the uncertainties of form
factors in the space-like region.


\begin{thebibliography}{10}

\bibitem{Nakamura:2010zzi}
K. Nakamura et~al.,
\newblock J. Phys. G37 (2010) 075021.

\bibitem{Ciulli:1975sm}
S. Ciulli, C. Pomponiu and I. Sabba-Stefanescu,
\newblock Phys.Repts.  (1975).

\bibitem{'tHooft:1973jz}
G. 't~Hooft,
\newblock Nucl. Phys. B72 (1974) 461.

\bibitem{Witten:1979kh}
E. Witten,
\newblock Nucl. Phys. B160 (1979) 57.


\bibitem{Arriola:2011en} 
  E.~Ruiz Arriola and W.~Broniowski,
\newblock  Contribution to Bled Workshops in Physics. Vol. 12 No. 1. 
arXiv:1110.2863 [hep-ph].


\bibitem{Kang:1975cq}
J. Kang and H.J. Schnitzer,
\newblock Phys.Rev. D12 (1975) 841.

\bibitem{RevModPhys.34.394}
G.F. Chew,
\newblock Rev. Mod. Phys. 34 (1962) 394.

\bibitem{Anisovich:2000kx}
A.V. Anisovich, V.V. Anisovich and A.V. Sarantsev,
\newblock Phys. Rev. D62 (2000) 051502, hep-ph/0003113.

\bibitem{RuizArriola:2010fj}
E. Ruiz~Arriola and W. Broniowski,
\newblock Phys.Rev. D81 (2010) 054009, 1001.1636.

\bibitem{Masjuan:2012gc}
P. Masjuan, E.Ruiz~Arriola and W. Broniowski,
\newblock Phys.Rev. D85 (2012) 094006, 1203.4782.

\bibitem{Afonin:2006wt}
S. Afonin,
\newblock Eur.Phys.J. A29 (2006) 327, hep-ph/0606310.

\bibitem{Glozman:2007ek}
L.Y. Glozman,
\newblock Phys.Rept. 444 (2007) 1, hep-ph/0701081.

\bibitem{deTeramond:2008ht}
G.F. de~Teramond and S.J. Brodsky,
\newblock Phys.Rev.Lett. 102 (2009) 081601, 0809.4899.

\bibitem{Ebert:2009ub}
D. Ebert, R. Faustov and V. Galkin,
\newblock Phys.Rev. D79 (2009) 114029, 0903.5183.

\bibitem{Broniowski:2000bj}
W. Broniowski and W. Florkowski,
\newblock Phys.Lett. B490 (2000) 223, hep-ph/0004104.

\bibitem{Broniowski:2004yh}
W. Broniowski, W. Florkowski and L.Y. Glozman,
\newblock Phys.Rev. D70 (2004) 117503, hep-ph/0407290.

\bibitem{Cohen:2011cr}
T.D. Cohen and V. Krejcirik,
\newblock J.Phys.G G39 (2012) 055001, 1107.2130.

\bibitem{Borsanyi:2010bp}
Wuppertal-Budapest Collaboration, S. Borsanyi et~al.,
\newblock JHEP 1009 (2010) 073.

\bibitem{Masjuan:2012sk}
P. Masjuan, E. Ruiz~Arriola and W. Broniowski,
\newblock (2012), 1210.0760.

\end{thebibliography}

\end{document}